\documentclass{article} 
\usepackage{iclr2027_conference,times}

\usepackage{amsmath,amsfonts,bm}

\def\eqref#1{equation~\ref{#1}}

\def\1{\bm{1}}

\DeclareMathAlphabet{\mathsfit}{\encodingdefault}{\sfdefault}{m}{sl}
\SetMathAlphabet{\mathsfit}{bold}{\encodingdefault}{\sfdefault}{bx}{n}

\usepackage{hyperref}
\usepackage{url}
\usepackage{graphicx}
\usepackage{booktabs}
\usepackage{longtable}
\usepackage{enumitem}
\usepackage{capt-of}
\usepackage{xcolor}
\usepackage{latexml}

\iflatexml
\else
  \usepackage[most]{tcolorbox}
\fi

\usepackage{listings}
\usepackage{pifont}
\lstdefinelanguage{prompt}{}
\iflatexml
  \newenvironment{promptbox}[3]{%
    \par\medskip
    \noindent\textbf{#1}\par
  }{%
    \par\medskip
  }
\else
  \newenvironment{promptbox}[3]{%
    \begin{tcolorbox}[
      title={#1},
      colback=#2,
      colframe=#3,
      coltitle=white,
      enhanced,
      breakable,
      fonttitle=\bfseries
    ]%
  }{%
    \end{tcolorbox}%
  }
\fi

\title{Discovery-Probability Graph Planning for Embodied Agents in Tasks with Hidden Cooperation Requirements}

\author{
\textbf{Zihuang Cai}\textsuperscript{1},
\textbf{Changjian Chen}\textsuperscript{1,\ensuremath{\dagger}},
\textbf{Zhuo Tang}\textsuperscript{1,\ensuremath{\dagger}},
\textbf{Chang Tang}\textsuperscript{2},
\textbf{Kenli Li}\textsuperscript{1}
\\
\textsuperscript{1}College of Computer Science and Electronic Engineering,
Hunan University
\\
\textsuperscript{2}Geely Holding Group,
China
}

\iclrfinalcopy

\newcommand{\methodname}{DPGP}

\begin{document}

\maketitle

\fancyhead{}
\renewcommand{\headrulewidth}{0pt}

\begingroup
\renewcommand{\thefootnote}{\fnsymbol{footnote}}
\footnotetext[2]{Corresponding authors.}
\endgroup

\begin{abstract}

Embodied agents are increasingly capable of perceiving, planning, and acting autonomously in dynamic environments, demonstrating strong performance on complex cooperative tasks.
Recent advances typically assume that cooperative tasks are known in advance. In real-world applications, however, hidden-cooperation tasks are more common: the need for cooperation only becomes apparent after an agent observes and interacts with task-related targets.
In this paper, we first present HiddenCoop, a benchmark for effectively evaluating agent cooperation in hidden-cooperation tasks. 
We further propose Discovery-Probability Graph Planning (DPGP), a method for enabling effective agent cooperation in such tasks. 
At the core of DPGP is a discovery-probability graph that encodes the probability distribution over the number and locations of task-related targets, facilitating effective and efficient information exchange among agents. 
A search algorithm is then used to derive action plans that minimize expected costs.
Experiments show that DPGP achieves cooperative-task success rates of 99.3\% on C-WAH and 58.3\% on TDW-MAT, surpassing all compared baselines, while also reducing overall embodied execution time and substantially lowering LLM token usage.
Our code is available at \url{https://github.com/hnu-vis/DPGP}.

\end{abstract}
\section{Introduction}

Embodied agents show a steadily increasing capacity to perceive and interact with their environments in human-like ways, while autonomously handling planning, decision-making, and action execution~\citep{pmlr-v205-ichter23a,pmlr-v205-huang23c}.
These advances have generated considerable interest in such agents, particularly for tasks requiring cooperative execution in dynamic environments~\citep{seo2026pce}.
Consider two household agents rearranging a home: each may search different rooms and transport objects individually, yet through interaction they discover that some objects require both agents. 
Such tasks, whose collaborative requirements must be discovered rather than specified in advance, pose a significant challenge to \emph{agent cooperation planning}: the agents must share newly acquired evidence, coordinate a rendezvous, and revise the allocation of the remaining work.

\begin{figure}[!t]
    \centering
    \setlength{\abovecaptionskip}{8pt}
    \includegraphics[width=\linewidth]
    {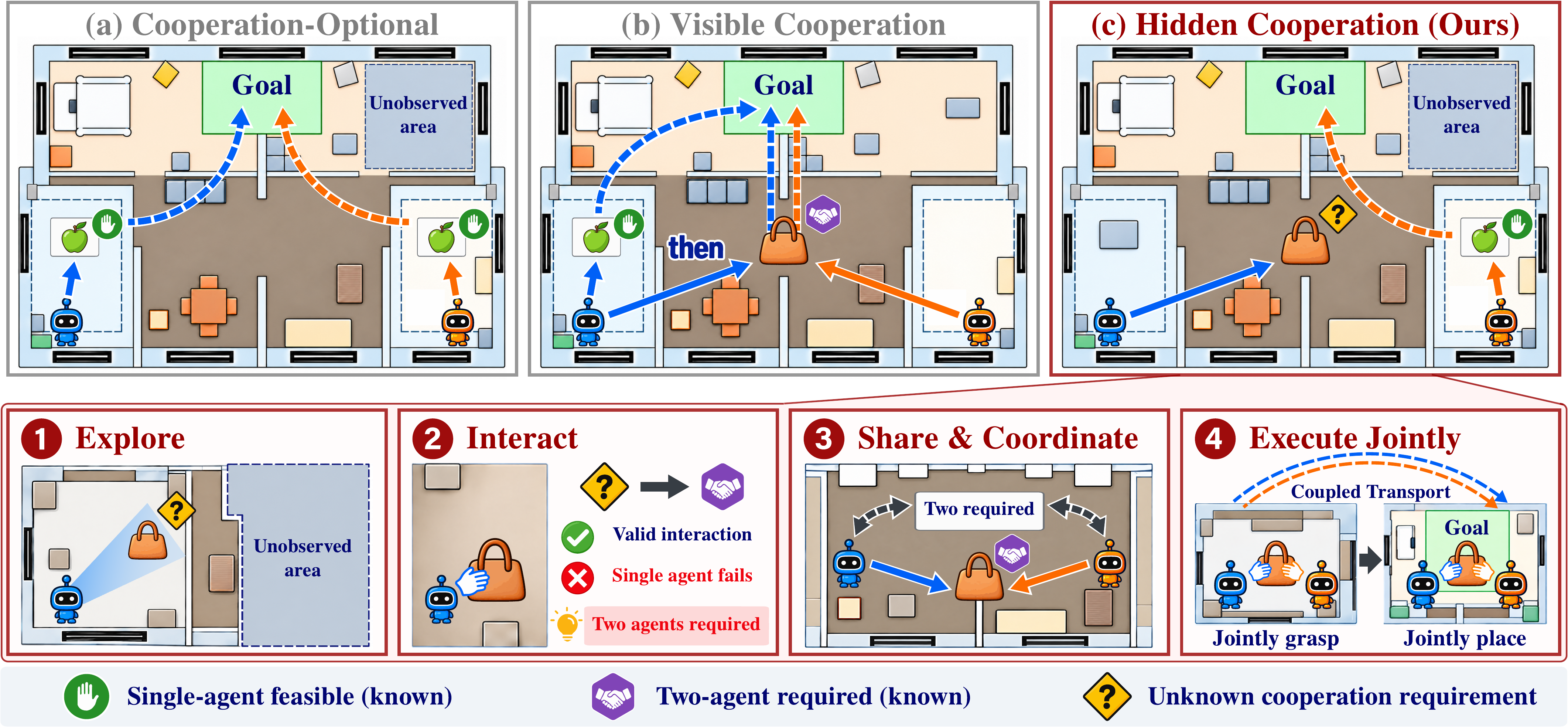}
    \caption{Representative agent cooperation settings. 
    (a) Cooperation-Optional Tasks, where each task can be handled independently by a single agent, but involving multiple agents can further improve results. 
    (b) Visible-Cooperation Tasks, where certain tasks inherently demand cooperation from multiple agents and this need is known in advance. 
    (c) Hidden-Cooperation Tasks, where the need for cooperation only becomes apparent after an agent interacts with task-related targets.}
    \label{fig:teaser}
\end{figure}

To tackle the challenges in agent cooperation planning, many methods have been proposed~\citep{zhang2024coela,liu2025capo,zu2025cots,seo2025reveca,jain2020cordial,pmlr-v306-kang26i}.
Early studies have primarily focused on \emph{Cooperation-Optional Tasks} (Figure~\ref{fig:teaser}(a)), in which each task can be handled independently by a single agent, but involving multiple agents can further improve results through parallel exploration and task allocation~\citep{zhang2024coela,chang2025partnr,liu2025capo,zu2025cots,seo2025reveca}.
However, such methods typically execute a task immediately upon discovering it, without considering whether to recruit other agents for joint execution.
More recent work has begun to investigate \emph{Visible-Cooperation Tasks} (Figure~\ref{fig:teaser}(b)), where certain tasks inherently demand multi-agent cooperation~\citep{jain2020cordial,pmlr-v306-kang26i,mandi2024roco,nayak2024llamar}.
However, these methods generally assume that these cooperation requirements are known in advance.
In practice, \emph{Hidden-Cooperation Tasks} (Figure~\ref{fig:teaser}(c)) are more common: the need to cooperate only becomes apparent after an agent observes or interacts with task-related targets.
In such tasks, the agents must first \underline{explore} the environment to establish situational awareness (Figure~\ref{fig:teaser}\ding{172}); then \underline{interact} with task-related targets to identify those requiring cooperation (Figure~\ref{fig:teaser}\ding{173}); next \underline{exchange} the information gathered and align plans with others (Figure~\ref{fig:teaser}\ding{174}); finally \underline{execute jointly} the hidden-cooperation tasks (Figure~\ref{fig:teaser}\ding{175}).
This complex process reveals a critical yet underexplored problem: \textbf{how to evaluate and enable effective agent cooperation in hidden-cooperation tasks when cooperation requirements must be discovered by the agents?}

In this paper, we introduce HiddenCoop, a benchmark for evaluating agent cooperation in hidden-cooperation tasks, and \emph{Discovery-Probability Graph Planning} (\methodname{}), a method for enabling effective agent cooperation in such tasks.
HiddenCoop is built on C-WAH and TDW-MAT~\citep{zhang2024coela} and includes both ordinary and cooperative household tasks. 
Cooperative tasks can only be completed by two agents, a requirement that remains unknown to the agents until they attempt them.
To model the uncertainty during discovering and completing these cooperative tasks, \methodname{} builds its core around a discovery-probability graph.
This graph encodes, for a given cooperative task, the probability distribution over the number of task-related targets and their locations.
As discoveries unfold, the graph is updated dynamically and shared among the agents, enabling precise information sharing and efficient transmission. 
Based on this graph, 
a search algorithm is applied to find the action plans that minimize the expected costs.
Experiments show that \methodname{} achieves cooperative-task success rates of 99.3\% on C-WAH and 58.3\% on TDW-MAT, exceeding all compared baselines.
It also achieves lower overall embodied execution time and substantially lower LLM token usage compared to baselines.
In summary, our contributions are threefold:

\begin{itemize}[leftmargin=*, labelindent=0pt]
    \item \textbf{A new benchmark, HiddenCoop}, for evaluating agent cooperation in hidden-cooperation tasks. 
    \item \textbf{A Discovery-Probability Graph Planning method} for enabling effective agent cooperation in hidden-cooperation tasks.
    \item \textbf{A series of experimental results} demonstrating the effectiveness and efficiency of our method.

\end{itemize}

\section{Related Work}

\subsection{Embodied Multi-Agent Benchmarks}

\paragraph{Tasks with individually executable subgoals.}
C-WAH and TDW-MAT study household exploration, information sharing, and task allocation under partial observability; PARTNR additionally introduces spatial, temporal, and capability constraints \citep{zhang2024coela,chang2025partnr}.
In these settings, an appropriately capable agent can execute each object-handling operation without mandatory joint handling of the same object.
HiddenCoop retains household exploration and multiple goals while introducing object-handling operations that require both agents.

\paragraph{Tasks with mandatory joint object handling.}
FurnMove and the large-object transport scenarios in Moving Out require sustained joint handling \citep{jain2020cordial,pmlr-v306-kang26i}.
Cooperation requirements depend on observed object properties or capability limits, as in macro-action warehouse tasks and GroundAct \citep{amato2019modeling,wang2025groundact}.
TIHDP and proactive collaboration methods also use interaction or execution feedback to assess feasibility and assistance needs \citep{naito2025tihdp,li2026proactive}.
Building on these precedents, HiddenCoop specifies an explicit information protocol: an unconfirmed category's requirement remains hidden to ordinary observation and is revealed through valid grasp interactions.
Evidence is initially available to participating agents, and confirmed requirements are reusable across same-category instances within the episode.

\subsection{Embodied Multi-Agent Planning}

\paragraph{Planning for individually executable subgoals.}
CoELA, REVECA, and CaPo use communication and task information to coordinate execution and revise plans \citep{zhang2024coela,seo2025reveca,liu2025capo}.
CoTS searches collaborative plans, whereas COMBO uses a compositional world model to simulate joint-action outcomes \citep{zu2025cots,zhang2025combo}.
PCE evaluates uncertain scenarios through likelihood, goal-directed gain, and execution cost, while CoBel-World maintains structured beliefs \citep{seo2026pce,wang2025cobelworld}.
KGLAMP connects knowledge graphs to symbolic planning and evaluates repairs by likelihood and plan cost \citep{shek2026kglamp}.
\methodname{} focuses on task-constrained resource-discovery distributions, comparing exploration and transport candidates on shared sampled outcomes under explicit joint execution constraints.

\paragraph{Planning for mandatory joint object handling.}
Macro-action decentralized policies coordinate joint execution through temporally extended actions, while CT-TCBS jointly plans team formation and transport using specified task locations and team-size requirements \citep{amato2019modeling,zhou2026cttapf}.
Learned policies and interaction feedback also support adaptive assistance \citep{naito2025tihdp,li2026proactive}.
RoCo combines language-based coordination with motion planning and feedback \citep{mandi2024roco}, while LLaMAR applies feedback-driven joint planning to tasks including rescue with predefined two-agent carrying requirements \citep{nayak2024llamar}.
\methodname{} focuses on the choice between further exploration and currently executable work, accounting for object assignments and both agents' ongoing actions.
Plans are revised using local observations and communicated evidence, incorporating newly confirmed cooperation requirements.
\section{Benchmark}

HiddenCoop extends C-WAH and TDW-MAT \citep{zhang2024coela} with hidden cooperation requirements. The resulting tasks combine individually executable subgoals with mandatory joint object handling under partial observability and decentralized control.

\subsection{Problem Formulation}

Two agents, indexed by \(i\in\{0,1\}\), work toward a public goal specification \(G\), which identifies target categories, required counts, and destination conditions.
At decision step \(t\), each agent acts using only its history of local observations and actions, execution feedback, and messages delivered to it.
The action and communication rules are public, but complete object locations and unconfirmed cooperation requirements are not provided.

For task-relevant category \(c\), we distinguish its requirement from agent \(i\)'s confirmed knowledge:
\begin{equation}
    \rho(c)\in\{1,2\}, \qquad
    \kappa_t^i(c)\in\{\mathrm{unknown},1,2\}.
    \label{eq:requirement}
\end{equation}
Here, \(\rho(c)=1\) permits individual handling subject to other action preconditions, whereas \(\rho(c)=2\) requires both agents.
Requirements are fixed within each episode and shared across task-relevant instances of a category; \(\mathrm{unknown}\) denotes unconfirmed knowledge, not a third object type.

Ordinary observation does not reveal unconfirmed requirements.
Confirmed category knowledge applies to other instances within the episode but reveals neither unseen objects nor their locations and is not automatically shared with the partner.
Agents must complete all goals within an execution-time budget accounting for the costs of both physical actions and communication.

\subsection{Benchmark Construction}

\paragraph{Requirement discovery and joint execution.}
Solo or paired grasp attempts reveal requirements to participating agents only after passing public preconditions independent of the hidden attribute, including reachability and occupancy checks.
A solo attempt on a cooperative object is then blocked, whereas a paired attempt on an ordinary object proceeds as an individual grasp.
Failed precondition checks do not reveal requirements: discovery requires valid interaction, not necessarily a failed solo grasp.
Cooperative objects require paired grasping and release, spatial coupling during transport, and continuous occupancy of both agents, preventing independent handling of other objects.
These task-level pairing, occupancy, and distance constraints do not explicitly model shared-load dynamics.
Successful execution is verified from the resulting simulator state rather than action submission.

\paragraph{Environment instantiation and matched tasks.}
C-WAH provides room-level symbolic observations, with closed-container contents hidden until opened.
Goals specify category counts and \texttt{ON}/\texttt{INSIDE} relations and accept eligible same-category instances; these counts do not specify the total eligible supply available in the scene.
TDW-MAT provides local visual observations with ground-truth instance masks for object identification.
We use a container-free transport setting that excludes initially contained targets.
Active-target category counts are public, but locations must be acquired through local observation or communication.
Every active target must be transported and released in the goal region; same-category instances are not interchangeable.

Each source task yields matched ordinary and cooperative episodes with identical initial scenes, agent positions, and semantic goals.
Ordinary episodes permit individual handling of all task-relevant objects, while cooperative variants require both agents for selected categories.
Both variants share the same observation and action interfaces, the communication protocol, and the execution-time model; episode-variant labels remain hidden.
C-WAH contains 50 matched pairs (100 episodes) and TDW-MAT 24 (48 episodes), supporting controlled comparisons of the effects of hidden cooperation requirements.
Full execution protocols are provided in Appendix~\ref{app:benchmark}.
\section{Method}
\label{sec:method}

\begin{figure}[!t]
    \centering
    \includegraphics[width=\linewidth]{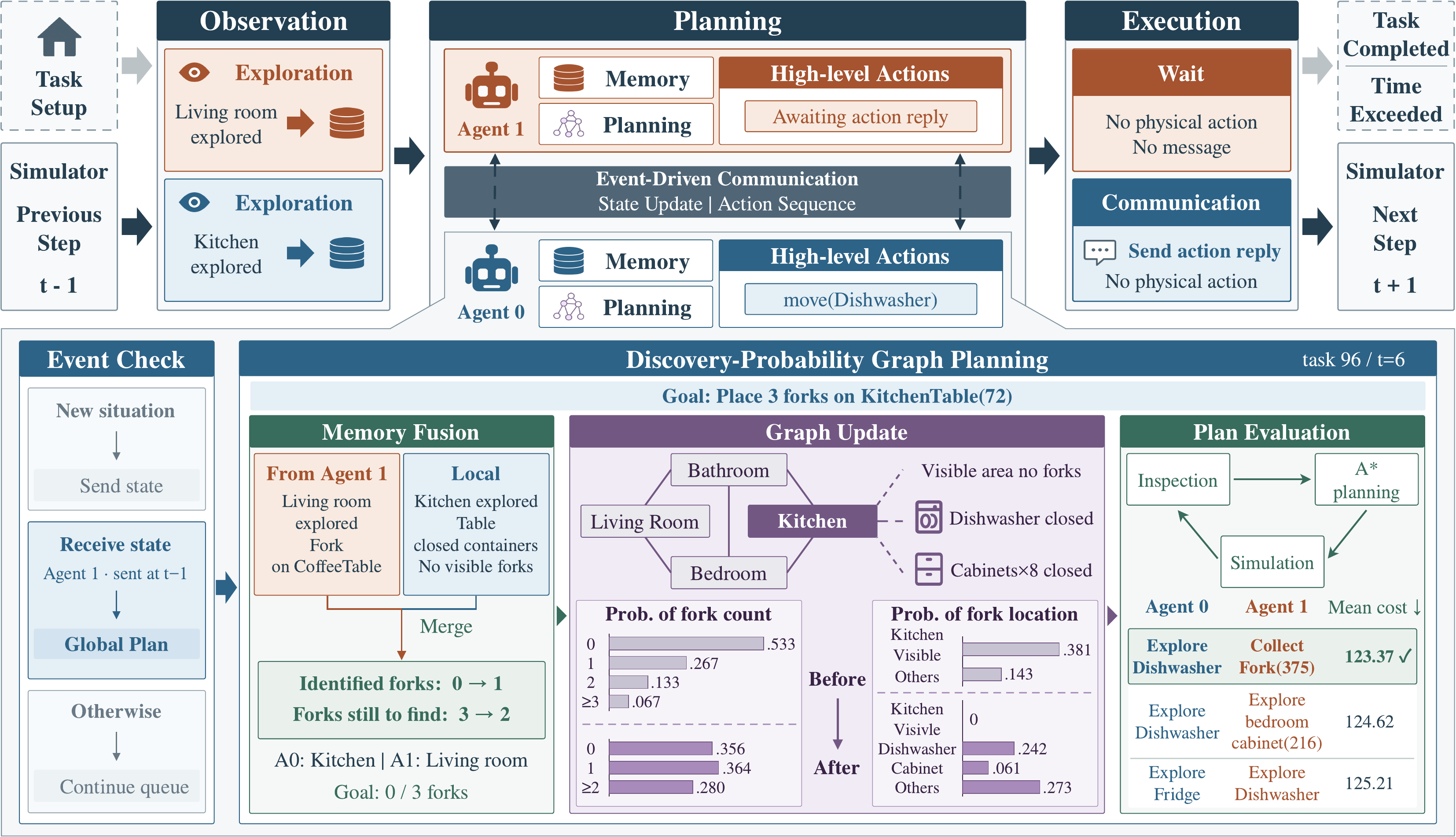}
    \caption{Overview of \methodname{}, showing the time-step execution loop and a step example.}
    \label{fig:method}
\end{figure}

Discovery-Probability Graph Planning (\methodname{}) coordinates exploration and transport using confirmed knowledge and probabilistic resource-discovery beliefs.
As shown in Figure~\ref{fig:method}, each agent follows a time-step loop: receiving local observations, execution feedback, and delivered messages, updating persistent memory, then planning and executing.
New observations and feedback close the loop.
We first outline the framework, then detail discovery-probability planning.

\subsection{Framework Overview}
\label{sec:framework}

Our framework consists of three modules:
(1) an Observation Module that maintains information across time steps; (2) a Planning Module that coordinates joint exploration and transport; and (3) an Execution Module that translates assigned plans into environment actions.

\paragraph{Observation Module.}
Each agent $i$ accumulates information from local observations, execution feedback, and delivered messages.
Its confirmed knowledge is represented by a fact graph $\mathcal F_t^i$, recording public goals, observed entities and spatial relations, confirmed cooperation requirements, exploration evidence, and known agent states.
A separate discovery belief $\mathcal B_t^i$ represents uncertainty about the counts and locations of undiscovered task-relevant resources.
The fact graph preserves available evidence, while the belief supports reasoning about potential discoveries.
LLM hypotheses and sampled objects are never recorded as confirmed facts.

\paragraph{Planning Module.}
The module first checks whether coordination is needed.
Initialization, task-relevant discoveries, newly confirmed cooperation requirements, execution failures, and completion of the assigned actions can trigger replanning; otherwise, agents continue their valid actions.
The planning agent combines local and received facts and invokes \methodname{} to jointly plan exploration and transport.
Discovery probabilities characterize potential exploration outcomes, while deterministic joint search evaluates feasible continuations under known execution constraints.
Candidates are compared on shared sampled outcomes, producing action sequences for both agents.
The partner's assigned actions are communicated through the public channel, with messages entering the recipient's next-step observation.
Section~\ref{sec:planning} details memory fusion, graph update, and plan evaluation.

\paragraph{Execution Module.}
The module translates each agent's assigned actions into executable navigation and object-handling actions, maintaining ongoing actions across time steps.
Confirmed cooperation requirements constrain paired grasping, coupled transport, and release.
Objects with unconfirmed requirements are optimistically modeled as ordinary for planning only; their requirements remain unknown, and execution remains subject to environment checks.
Action outcomes are verified through subsequent observations or feedback rather than command submission.
Valid interactions can confirm cooperation requirements, updating memory and triggering replanning.

\subsection{Discovery-Probability Graph Planning}
\label{sec:planning}

At each coordination round, \methodname{} constructs a joint plan through three stages: (1) Memory Fusion integrates available evidence from observations and messages; (2) Graph Update incorporates confirmed structure and updates discovery beliefs; and (3) Plan Evaluation compares exploration and transport candidates through shared sampling and deterministic joint search.

\paragraph{Memory Fusion.}
The planning agent combines local facts with information received from its partner.
Objects are aligned by identity, and repeated observations are consolidated to avoid counting evidence twice.
Dynamic states are updated using evidence timestamps and sources.
Each agent's execution state comes from its own account: local memory for the planning agent and delivered messages for its partner.
The fused fact graph records known objects, spatial relations, exploration evidence, and execution constraints without introducing assumptions about unobserved facts.
For the current round, we denote the planner's fact graph and separate discovery belief by $\mathcal F$ and $\mathcal B$.

\paragraph{Graph Update.}
The fused facts determine known objects, spatial relations, and candidate search regions; the discovery belief models the resources that further exploration may reveal.
Hypothetical discoveries remain separate from confirmed facts.
For each task-relevant category $c$, let $d_c$ denote the remaining goal demand and $s_c$ the number of confirmed, unfinished objects that remain eligible to satisfy this demand.
The demand not covered by these known objects is
\begin{equation}
    R_c = \max(0,d_c-s_c).
    \label{eq:dpgp_gap}
\end{equation}
Only confirmed, unfinished objects eligible for the remaining goals contribute to $s_c$; previously identified objects are excluded from the undiscovered supply modeled below.
This avoids counting known identities twice.
Let $U_c$ denote the unknown number of undiscovered objects eligible to satisfy the remaining category-$c$ demand.
We represent this quantity using a task-capped random variable $K_c=\min(U_c,R_c)$ and denote its probability mass function by $\pi_c$:
\begin{equation}
    \pi_c(k)=\Pr(K_c=k),
    \qquad k\in\{0,\ldots,R_c\}.
    \label{eq:dpgp_count}
\end{equation}
All probabilities are conditioned on the information currently available to the planner.
For $R_c>0$, the final bin aggregates all cases with at least $R_c$ undiscovered eligible objects. For planning, we sample at most $R_c$ undiscovered objects, since additional instances exceed the current demand gap.
When $R_c=0$, no additional undiscovered supply is modeled.
Public task information may determine how many eligible objects remain undiscovered. In this case, $\pi_c$ assigns probability one to the corresponding capped count, while their locations may still be uncertain.

To model potential object locations, let $\mathcal R_c$ denote a set of mutually exclusive candidate regions.
For each region $r\in\mathcal R_c$, let $w_{cr}$ denote the probability that an undiscovered category-$c$ object is located there under the model, with $\sum_{r\in\mathcal R_c}w_{cr}=1$.
Given $K_c=k$, we independently sample a region for each of the $k$ objects from this distribution.
Let $N_{cr}$ denote the number of sampled objects assigned to region $r$. The modeled probability of discovering at least one such object in that region is
\begin{equation}
    \Pr(N_{cr}>0)
    =1-\sum_{k=0}^{R_c}\pi_c(k)(1-w_{cr})^k.
    \label{eq:dpgp_discovery}
\end{equation}
This estimate uses the task-capped count model rather than the full scene inventory.
Regional counts share the same sampled quantity $K_c$, so discoveries across regions are evaluated jointly rather than as independent events.
Given the available evidence, we model and sample object counts and locations separately for each category, without explicitly modeling dependencies between categories.

We initialize discovery distributions over uncertain object counts and locations using LLM-derived semantic priors.
LLMs have been used to construct world models and beliefs for planning under uncertainty \citep{zhao2023llmmcts,tang2025trupomdp}.
Here, the LLM assesses semantic plausibility, while explicit computations normalize its outputs into probabilities and update the distributions from observed evidence.
For a finite hypothesis set $\mathcal H$ consistent with public information and confirmed facts, the LLM assigns a support level $\ell_h\in\{0,1,2,3,4\}$ to each hypothesis $h\in\mathcal H$.
Each hypothesis specifies a count or a semantic region group.
The prior probability of $h$ is
\begin{equation}
    p_0(h)=
    \frac{2^{\ell_h}}
    {\sum_{h'\in\mathcal H}2^{\ell_{h'}}}.
    \label{eq:dpgp_prior}
\end{equation}
The prior initializes $\pi_c$ only when the undiscovered count is uncertain; otherwise, public counts and confirmed discoveries determine its point mass.
For locations, the LLM scores semantic groups of candidate regions, and each group's probability is divided equally among its members to initialize $w_{cr}$.
This prevents larger groups from receiving more mass solely because they contain more regions.
These probabilities express semantic plausibility, not empirically calibrated frequencies.

Confirmed discoveries update known supply and the demand gap; publicly determined counts are updated from the inventory.
For uncertain counts, suppose a complete inspection of region $r$ reveals $y$ previously undiscovered eligible category-$c$ objects located there.
Using pre-inspection values of $R_c$, $\pi_c$, and $w_{cr}$, set $y^\star=\min(y,R_c)$.
Let $\pi_c^+(m)$ denote the updated probability of $m$ objects remaining undiscovered under the task-capped model.
Then
\begin{equation}
    \pi_c^+(m)
    \propto
    \pi_c(m+y^\star)
    \binom{m+y^\star}{y^\star}
    w_{cr}^{y^\star}(1-w_{cr})^m,
    \quad
    m=0,\ldots,R_c-y^\star,
    \label{eq:dpgp_update}
\end{equation}
with normalization over $m$ when the total weight is positive.
The update treats the tail event $U_c\geq R_c$ as exactly $R_c$ undiscovered objects, yielding an approximate rather than exact posterior over the full inventory.
Completely inspected regions are removed and remaining location weights renormalized when positive mass remains; zero-likelihood evidence triggers belief repair.
Evidence and identities are incorporated only once, and confirmed objects remain in the fact graph regardless of the cap.
Partial observations update the inventory but do not justify excluding an entire region.
Environment-specific updates and repair procedures are detailed in Appendix~\ref{app:method}.

\paragraph{Plan Evaluation.}
Given the current fact graph $\mathcal F$ and discovery belief $\mathcal B$, the planner constructs a bounded candidate set $\mathcal U$.
Each $u\in\mathcal U$ assigns the agents to known-object transport, parallel exploration, or a combination of exploration and transport.
Candidates respect current holdings, ongoing actions, and cooperative occupancy.
Exploration screening balances expected goal contribution against navigation cost: interchangeable-object tasks consider unmet demand and potential delivery savings, whereas all-target tasks consider the expected number of undiscovered targets.
Feasible known-object transport alternatives are retained.

All candidates are evaluated on the same $M$ discovery outcomes $\{\omega^{(j)}\}_{j=1}^{M}$ sampled from $\mathcal B$, specifying object counts and locations.
Deterministic joint A* search plans transport over both agents' positions, holdings, ongoing actions, and remaining goals.
Transitions enforce action preconditions, carrying capacity, object exclusivity, and confirmed cooperation requirements.
For exploration candidates, simulated inspections reveal sampled objects before transport planning.
Only known objects and those revealed in simulation enter the search; unrevealed samples remain unavailable.
Sampled discoveries inform evaluation but never become object references for real actions.

Let $J(u,\omega^{(j)})$ denote the total planning cost of candidate $u$ under outcome $\omega^{(j)}$, including the costs of its executed action and simulated continuation, plus a penalty for unresolved demand.
This penalty prevents unfinished work from yielding artificially low costs.
The estimated planning cost is
\begin{equation}
    \widehat J(u)
    =
    \frac{1}{M}\sum_{j=1}^{M}J(u,\omega^{(j)}).
    \label{eq:dpgp_objective}
\end{equation}
Costs are environment-specific planning proxies, not measured execution time; their parameters, continuation procedures, and residual penalties appear in Appendix~\ref{app:planning-cost}.
Selection combines these estimates with uncertainty and tie-breaking rules, requiring sufficient improvement over a feasible transport baseline where applicable (Appendix~\ref{app:candidate-evaluation}).
Selection is therefore not based on mean cost alone.
Only the selected actions grounded in available facts are executed; sampled discoveries remain hypothetical.
Actual observations and feedback update $\mathcal F$ and $\mathcal B$ for the next round.
\section{Experiment}
\label{sec:experiments}

\subsection{Experimental Settings}
\label{sec:experimental-settings}

\paragraph{Tasks and Baselines.}
We evaluate on 50 C-WAH and 24 TDW-MAT matched ordinary/cooperative task pairs under the public HiddenCoop interfaces.
Each method is evaluated with three runs per C-WAH task and one per TDW-MAT task, yielding 300 and 48 episode records, respectively, with matched task and seed schedules across methods.
Trajectories are collected with execution budgets of 600 s for C-WAH and 20 s for TDW-MAT, without additional step or frame limits. The main results use evaluation budgets of 200 s and 20 s, respectively.
We compare \methodname{} with CoELA, CoTS, REVECA, and PCE \citep{zhang2024coela,zu2025cots,seo2025reveca,seo2026pce}.
All use GPT-4o-mini through environment-specific adapters, retaining method-specific prompting and sampling settings.
Interface adaptations are described in Appendix~\ref{app:baselines}, and the configuration policy is summarized in
Appendix~\ref{app:baseline-params}.

\paragraph{Evaluation Metrics.}
We report within-budget task success rate (Task SR), embodied execution time (EET), and time-normalized goal-completion area under the curve (TGC-AUC); formal definitions are given in Appendices~\ref{app:benchmark-time}, \ref{app:benchmark-success}, and~\ref{app:metric-definitions}.
EET includes physical execution, communication, and idle costs but excludes LLM inference and planning wall-clock time.
TGC-AUC integrates the completed-goal fraction over the evaluation budget and normalizes by that budget, rewarding earlier goal completion.
Results include successful and valid failed episodes, grouped into all, ordinary, and cooperative tasks.
Goal completion rate (Goal CR) pools completed over required goal instances; token usage sums both agents' recorded LLM usage per episode.

\paragraph{Additional Analyses.}
We reevaluate recorded execution trajectories at 100--600 seconds in 50-second increments for C-WAH and 10--20 seconds in one-second increments for TDW-MAT.
These are cutoff analyses, not budget-conditioned reruns; completion after each cutoff receives no success or AUC credit.
We also compare EET on the episodes completed successfully by all five methods.

\begin{table*}[t]
    \centering
    \caption{Main results comparing \methodname{} with baselines on HiddenCoop (C-WAH and TDW-MAT). The best results in each column are shown in \textbf{bold}, and the second-best is \underline{underlined}.}
    \label{tab:main_results}
    \small
    \setlength{\tabcolsep}{3.5pt}
    \renewcommand{\arraystretch}{1.08}
    \begin{tabular*}{\linewidth}{@{\extracolsep{\fill}}l*{10}{r}@{}}
        \toprule
        & \multicolumn{3}{c}{Task SR (\%) $\uparrow$}
        & \multicolumn{3}{c}{EET (s) $\downarrow$}
        & \multicolumn{3}{c}{TGC-AUC $\uparrow$}
        & Tokens (k) $\downarrow$ \\
        \cmidrule(lr){2-4}
        \cmidrule(lr){5-7}
        \cmidrule(lr){8-10}
        \cmidrule(l){11-11}
        Method
        & All & Ord. & Coop.
        & All & Ord. & Coop.
        & All & Ord. & Coop.
        & All \\
        \midrule
        \multicolumn{11}{@{}l}{
            \textbf{C-WAH}\quad\textit{Evaluation budget: 200 s}
        } \\
        \addlinespace[2pt]
        CoELA
        & \underline{91.7} & \underline{99.3} & \underline{84.0}
        & 125.54 & 106.76 & 144.31
        & 0.6267 & 0.6890 & \underline{0.5644}
        & 141.32 \\
        CoTS
        & 71.0 & \underline{99.3} & 42.7
        & 153.64 & 125.38 & 181.89
        & 0.4583 & 0.5898 & 0.3267
        & 139.71 \\
        REVECA
        & 84.0 & \textbf{100.0} & 68.0
        & 125.92 & 92.27 & 159.56
        & 0.5994 & 0.7267 & 0.4720
        & 254.62 \\
        PCE
        & 88.0 & \textbf{100.0} & 76.0
        & \underline{115.61} & \underline{90.07} & \underline{141.15}
        & \underline{0.6397} & \textbf{0.7407} & 0.5387
        & \underline{112.03} \\
        \methodname{} (ours)
        & \textbf{99.7} & \textbf{100.0} & \textbf{99.3}
        & \textbf{94.90} & \textbf{82.31} & \textbf{107.49}
        & \textbf{0.6874} & \underline{0.7270} & \textbf{0.6477}
        & \textbf{7.16} \\
        \midrule
        \multicolumn{11}{@{}l}{
            \textbf{TDW-MAT}\quad\textit{Budget: 20 s}
        } \\
        \addlinespace[2pt]
        CoELA
        & 43.8 & 62.5 & 25.0
        & 15.80 & 13.34 & 18.25
        & 0.4895 & 0.5986 & 0.3803
        & \underline{485.84} \\
        CoTS
        & 47.9 & \underline{70.8} & 25.0
        & 16.71 & 14.24 & 19.19
        & 0.4601 & 0.5691 & 0.3510
        & 2825.56 \\
        REVECA
        & \underline{52.1} & 66.7 & 37.5
        & 15.73 & 13.42 & 18.04
        & 0.5026 & 0.5895 & 0.4157
        & 1013.29 \\
        PCE
        & \textbf{64.6} & \textbf{83.3} & \underline{45.8}
        & \underline{14.58} & \underline{11.72} & \underline{17.44}
        & \underline{0.5452} & \underline{0.6302} & \underline{0.4601}
        & 1202.03 \\
        \methodname{} (ours)
        & \textbf{64.6} & \underline{70.8} & \textbf{58.3}
        & \textbf{12.61} & \textbf{10.78} & \textbf{14.45}
        & \textbf{0.6172} & \textbf{0.6722} & \textbf{0.5623}
        & \textbf{1.41} \\
        \bottomrule
    \end{tabular*}
\end{table*}

\begin{figure}[!t]
    \centering
    \includegraphics[width=\linewidth]{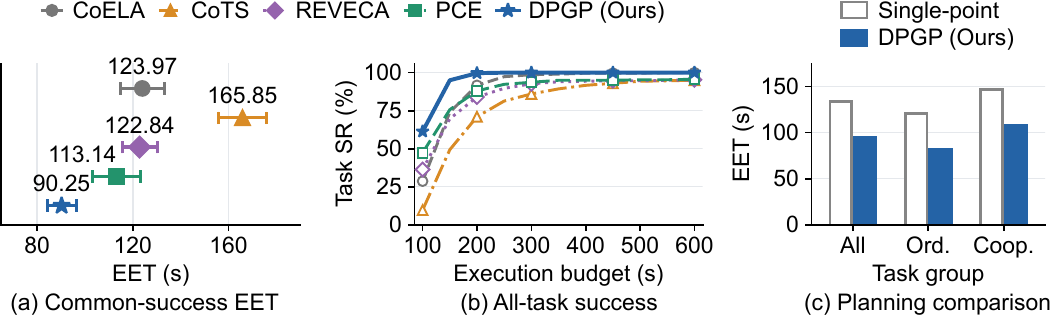}
    \caption{Additional analyses of execution efficiency and planning performance on C-WAH.}
    \label{fig:analysis}
\end{figure}

\subsection{Main Results}

Table~\ref{tab:main_results} shows that \methodname{} delivers the best overall and cooperative-task performance in both environments, achieving the highest Task SR, lowest EET, and highest TGC-AUC.
Its overall success rates reach 99.7\% on C-WAH and 64.6\% on TDW-MAT, while cooperative-task success reaches 99.3\% and 58.3\%, respectively.
Across all methods, cooperative tasks reduce success and delay goal completion, reflecting the combined burden of discovering cooperation requirements and executing joint actions.
\methodname{} is least affected by this transition, exhibiting the smallest ordinary-to-cooperative drops in Task SR and TGC-AUC, together with the smallest increase in EET, in both environments.
The framework also requires substantially fewer LLM tokens, averaging only 7.16k and 1.41k per evaluated episode on C-WAH and TDW-MAT, respectively.

Figure~\ref{fig:analysis} complements the main table by examining two effects of the execution budget.
First, failed episodes can consume the available budget, making average EET sensitive to the chosen cutoff.
Panel~(a) therefore compares the same 264 C-WAH records completed successfully by all five methods within the 600\,s collection budget.
On this common-success intersection, \methodname{} reduces EET by 20.2\% relative to the fastest baseline, PCE, demonstrating an efficiency advantage without failed-episode termination times.
Second, panel~(b) evaluates all C-WAH records across 100--600\,s cutoffs, placing the main table's 200\,s result in a broader budget range.
\methodname{} maintains the highest or tied-highest success rate throughout; at 150\,s, it achieves 95.0\% versus 75.7\% for the strongest baseline.
The advantage is therefore not confined to the selected cutoff, while convergence at larger budgets explains why final success alone can obscure execution-efficiency differences.

\subsection{Ablation Study}

We examine uncertainty-aware planning by comparing the full framework with a single-point prediction variant.
Both use LLM-derived semantic information to reason about unknown target locations.
The single-point variant commits to definite location hypotheses and performs deterministic joint search over the completed graph.
The full framework instead maintains location distributions, updates them with observations, and evaluates joint action candidates across shared sampled worlds.
This comparison targets the combined role of uncertainty representation and its use in planning.

On C-WAH, both variants achieve 100\% task success at the 600\,s collection budget, while the full framework reduces average EET from 133.98 to 95.04 seconds (29.1\%).
The improvements occur on both ordinary and cooperative tasks, as shown in Figure~\ref{fig:analysis}(c), indicating faster task progress and completion despite identical final success rates.

\subsection{Case Study}

\begin{figure}[!t]
    \centering
    \includegraphics[width=\linewidth]{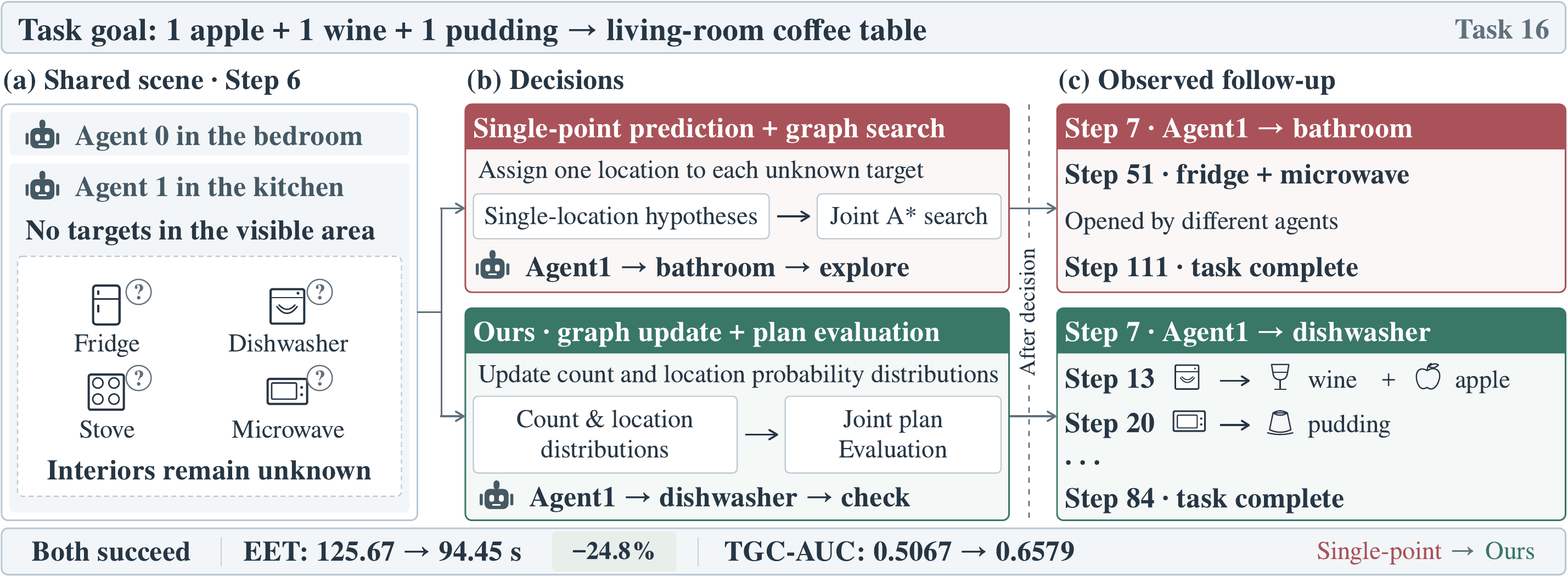}
    \caption{Search decisions of \methodname{} and the single-point variant at step 6 of C-WAH task 16.}
    \label{fig:casestudy}
\end{figure}

Figure~\ref{fig:casestudy} compares \methodname{} with the single-point variant on C-WAH task 16.
At step 6, no targets have been found in the kitchen's visible area, while the contents of newly observed containers remain unknown.
The single-point variant assigns one location to each unknown target and performs joint A* search, directing Agent 1 toward the bathroom.
In contrast, \methodname{} updates probability distributions over undiscovered target counts and locations and evaluates joint plans, directing Agent 1 to inspect the dishwasher.
Both runs succeed, but \methodname{} discovers the targets earlier, reducing EET from 125.67 to 94.45 seconds and increasing TGC-AUC from 0.5067 to 0.6579.
This case highlights the role of probabilistic belief updates and joint plan evaluation in adapting search decisions to newly observed information.
\section{Conclusion}

We introduced HiddenCoop to study multi-agent exploration and execution with interaction-revealed cooperation requirements.
Our framework, \methodname{}, combines confirmed facts with resource-discovery beliefs to evaluate exploration and transport jointly under execution constraints.
Experiments on C-WAH and TDW-MAT show strong cooperative-task performance, improved execution efficiency and goal progress, and substantially lower LLM token usage.
Common-success and execution trajectory analyses further distinguish execution efficiency from final task success.
These findings support evaluating potential discoveries together with their downstream execution costs when coordinating partially observable tasks.
Future work will investigate larger, heterogeneous teams and more complex cooperation requirements.

\section{AI use statement}
We used generative AI tools to assist with manuscript writing and polishing, including improving language clarity, coherence, and organization.
We also used
these tools for literature retrieval and discovery, including identifying relevant studies, summarizing prior work, and organizing the related-work discussion.
We take responsibility for the final content of this work, including text and claims produced with the aid of generative AI.

\bibliography{iclr2027_conference}
\bibliographystyle{iclr2027_conference}

\clearpage
\appendix
\section{Benchmark Construction}
\label{app:benchmark}

\subsection{Paired Tasks}
\label{app:benchmark-data}

HiddenCoop constructs an ordinary and a cooperative variant from each source task. Within a pair, the initial scene, goals, and agent starts are fixed; the cooperative variant additionally requires paired handling for selected target categories. C-WAH contains 50 pairs (100 tasks), and TDW-MAT contains 24 pairs (48 tasks). Table~\ref{tab:app-benchmark-composition} reports target instances across both variants. Each task has one shared goal set, so targets are counted once per task rather than once per agent.

Pairing keeps task content aligned when measuring the effect of hidden cooperation requirements. Dataset validators and evaluator-side target resolution check construction and scoring; they are not policies and do not supply privileged information to the evaluated methods.

\subsection{Information Available to Agents}
\label{app:benchmark-information}

Agents receive the public goals and their own admissible observations. In C-WAH, they observe the locally available scene graph, while the contents of closed containers remain hidden. In TDW-MAT, they receive local RGB-D observations, ground-truth segmentation for visible objects, and public room metadata; a local scan may still miss a target. A target's cooperative property becomes known only after a valid interaction reaches the hidden-attribute check. Invalid or out-of-range probes do not reveal it. The other agent obtains that information only through its own interaction or a delivered message.

Complete cooperative labels, full initial states, the complete runtime target-ID sets, global target locations, future states, and offline analysis annotations are excluded from prompts and online memory. Task-variant and source-pair identifiers are reserved for evaluation.

\subsection{Cooperative Handling}

At each decision step, an agent submits a physical-action field and an optional message. Under the evaluated implicit-pairing interface, matching ordinary grasp intents for the same object are interpreted as a paired grasp. During an active cooperative carry, matching release intents are interpreted as a paired release. Explicit cooperative actions are also supported. The environment checks the joint action but does not choose targets, summon a partner, or decide how the agents recover from failure.

Each agent can hold one target or occupy one cooperative role. A TDW-MAT grasp or drop may span multiple decision steps; an unfinished transaction must continue with \texttt{ongoing}. Pairing, distance, execution, and postcondition failures produce distinct public feedback. Rendezvous, recovery, and retry decisions remain the responsibility of the method.

\subsection{Communication and Embodied Time}
\label{app:benchmark-time}

In physical deployments, transmitting a short message does not occupy the same wall-clock interval as a physical manipulation; charging each message one full action step would overstate its cost. We therefore model communication as an exclusive channel whose duration scales with the delivered message length, while retaining mutual exclusion between sending a message and initiating a new physical action.

The evaluated setting uses exclusive-action communication: an agent sending a message cannot initiate a new physical action in that step, although its partner may act. Messages arrive in the next observation. Communication is charged by the number of Unicode codepoints actually delivered after truncation; suppressed or unsent content is not charged. The rates are 2,500 codepoints/s for C-WAH and 125,000 codepoints/s for TDW-MAT, and the maximum message length is 12,000 codepoints in both environments.

We measure cost in standardized seconds rather than decision steps or simulator frames because action durations differ substantially across action types and environments: a turn, opening a container, and a long displacement do not consume comparable time, so counting them as equal steps would misrepresent execution cost.

Let $P_t$ denote joint physical duration, $S_t=\max_a \ell_{t,a}/r$ the exposed message duration at rate $r$, and $I_t$ the minimum charge applied only to an otherwise idle step. Embodied execution time (EET) is
\begin{equation}
T_t=\max\{P_t,S_t,I_t\},\qquad
\mathrm{EET}=\sum_t T_t.
\label{eq:app-eet}
\end{equation}
C-WAH uses a standardized 5\,Hz time scale with fixed action charges applied identically to all evaluated methods. Table~\ref{tab:app-action-charges} reports duration statistics from verified successful executions under the environment's original joint-script mode as an empirical reference alongside these charges.

\begin{table}[t]
\centering
\caption{Benchmark composition. The ordinary and cooperative variants of each source task have the same number of target instances.}
\label{tab:app-benchmark-composition}
\small
\setlength{\tabcolsep}{4pt}
\begin{tabular}{@{}lrr@{}}
\toprule
 & C-WAH & TDW-MAT \\
\midrule
Tasks / source pairs & 100 / 50 & 48 / 24 \\
Ordinary / cooperative tasks & 50 / 50 & 24 / 24 \\
Targets in ordinary / cooperative variants & 209 / 209 & 103 / 103 \\
Ordinary / cooperative target instances & 304 / 114 & 160 / 46 \\
Total target instances & 418 & 206 \\
\bottomrule
\end{tabular}
\end{table}

\begin{table}[t]
\centering
\caption{Key conditions for paired handling. C-WAH uses its native \textsc{Close} relation for interactions; TDW-MAT uses metric distances.}
\label{tab:app-coop-lifecycle}
\small
\setlength{\tabcolsep}{3pt}
\renewcommand{\arraystretch}{1.08}
\begin{tabular}{@{}p{.16\columnwidth}p{.39\columnwidth}p{.39\columnwidth}@{}}
\toprule
Stage & C-WAH & TDW-MAT \\
\midrule
Grasp
& Same object; holder \textsc{Close} to it, supporter \textsc{Close} to holder or object.
& Same object; both agents within 2\,m of it and each other, with free capacity. \\
Carry
& One physical holder and one occupied supporter; soft/hard separation thresholds of 2/3 scene units.
& One holder and one occupied supporter; soft/hard thresholds of 2/4\,m. \\
Release
& Matching object, destination, and placement type; actual release and \textsc{On}/\textsc{Inside} relation are verified.
& Matching drop intents and valid roles and distance; delivery is checked after physical release. \\
\bottomrule
\end{tabular}
\end{table}

\begin{table}[t]
\centering
\caption{C-WAH action-duration statistics and standardized charges. Duration statistics are computed over verified successful segments from scripted executions under the original pre-compiled joint-script mode and provide an empirical reference for the fixed charges applied identically to all methods. Navigation is charged by horizontal displacement; its measured per-segment mean speed is 0.97 scene units/s (median 0.85). Turn yields no qualifying segments and is fixed by convention.}
\label{tab:app-action-charges}
\small
\setlength{\tabcolsep}{4pt}
\begin{tabular}{@{}lrrrrr@{}}
\toprule
Action & Segments & Mean (s) & Median (s) & p05--p95 (s) & Charge (s) \\
\midrule
Grasp    & 5{,}521 & 1.77 & 1.40 & 1.40--4.40 & 1.6 \\
Open     & 3{,}479 & 3.60 & 3.00 & 1.80--6.60 & 3.2 \\
Put in   &   995   & 1.67 & 1.40 & 1.40--4.20 & 2.8 \\
Put back & 1{,}635 & 3.10 & 1.60 & 1.40--6.00 & 2.8 \\
Navigation & 165{,}973 & -- & -- & -- & 1 unit/s \\
Turn     & -- & -- & -- & -- & 1.0 \\
\bottomrule
\end{tabular}
\end{table}

\subsection{Success and Budget Boundaries}
\label{app:benchmark-success}

C-WAH evaluates the required \textsc{On}/\textsc{Inside} relations in the current scene graph, so progress can decrease if a placed object is removed. TDW-MAT marks a target as delivered once it is unheld and within 3\,m of the designated bedroom bed; this delivery remains satisfied. Completion is checked after a full step. Completion exactly at budget $B$ counts, whereas completion after $B$ does not. The first step crossing the budget remains in the trace, but goals newly completed during that step are excluded from budget-limited success and progress.

For $G$ required target instances and $g(t)$ instances satisfied at $t$, time-weighted goal completion is
\begin{equation}
\mathrm{TGC\mbox{-}AUC}
=\frac{1}{B}\int_0^B\frac{g(t)}{G}\,dt.
\label{eq:app-tgc-auc}
\end{equation}
Progress is constant between completed steps. If an episode ends before $B$, its final valid progress is extended to the budget boundary. Task success, terminal goal completion, and termination reason are retained as separate quantities.
\section{Method Details}
\label{app:method}

This section details the belief representation, joint candidate evaluation, planning costs, and execution interface of \methodname{}. The two environments share the decision procedure but differ in what an inspection can establish and in the units used by their planners.

\subsection{Belief Representation and Evidence Updates}
\label{app:belief}

Each agent separates confirmed facts from hypotheses about unobserved targets. Object identities and relations, locations, cooperative properties, and execution states become facts through admissible observations, action feedback, or delivered messages. Sampled locations remain planning hypotheses until an actual observation confirms them. The language model supplies semantic location support; it does not determine a target's hidden cooperative property.

\paragraph{C-WAH.}
For category $c$, let $d_c$ be the remaining demand and $s_c$ the known, non-duplicated usable supply. The anonymous deficit is $R_c=\max(0,d_c-s_c)$. The count variable $K_c\in\{0,\ldots,R_c\}$ is capped at the task deficit: its last bin represents at least $R_c$ useful anonymous objects, of which at most $R_c$ need to be instantiated in planning. Known identities with unresolved locations are not counted again as anonymous discoveries. Such identities, and required destinations whose locations remain unknown, have separate categorical location variables.

Location alternatives include uninspected visible areas, hidden or residual structure, and real uninspected containers. Hidden and residual structure estimate future search work; executable inspection actions refer only to actual observed rooms or containers. A complete inspection conditions the anonymous count on the number found and removes the inspected region from the remaining alternatives. Inspecting a room's visible area does not establish that its closed containers are empty.

\paragraph{TDW-MAT.}
Public goal counts determine the number of required target instances. Unresolved slots represent required instances not yet bound to observed identities, completed, or held; there is no C-WAH-style count distribution. Additional objects of the same category cannot substitute for active targets. Location alternatives are real rooms, and an unresolved instance is sampled from its room distribution. Because a local scan can miss an object, an empty scan does not eliminate that room. Only the first genuine observation of an object's original, unmoved location supplies discovery-location evidence.

\paragraph{Semantic support.}
For a count or location row, the language model assigns each candidate an integer level $L_j\in\{0,1,2,3,4\}$. The method converts these levels into relative weights:
\begin{equation}
w_j=\frac{2^{L_j}}{\sum_{j'}2^{L_{j'}}}.
\label{eq:app-support-normalization}
\end{equation}
Level zero therefore retains positive support. These normalized values encode semantic plausibility; they have not been established as empirically calibrated probabilities. In C-WAH, containers of the same type in the same room share one semantic group's mass, which is divided among its actual members. Treating category roots as conditionally independent and allocating anonymous representatives independently within a root are modeling approximations.

For a C-WAH region $r$ with allocation weight $w_{c,r}$, the modeled probability of finding at least one useful anonymous object is
\begin{equation}
\Pr(N_{c,r}>0)
=1-\sum_{k=0}^{R_c}\Pr(K_c=k)(1-w_{c,r})^k.
\label{eq:app-region-discovery}
\end{equation}
Thus a location weight alone does not imply that an object exists there. Complete C-WAH inspections also update semantic-group experience. For qualifying positive and empty observations $h_{g,c}$ and $e_{g,c}$, respectively, the correction derived from a $\mathrm{Beta}(1,1)$ starting model is
$q_{g,c}=2(1+h_{g,c})/(2+h_{g,c}+e_{g,c})$.
Each observation contributes once, even if it persists across control steps. When new structure changes the alternatives, the conditioned count distribution is retained while missing location support is requested again. Confirmed facts take precedence over conflicting predictions, and affected belief components are repaired before a validated update is committed.

\subsection{Joint Candidate Evaluation}
\label{app:candidate-evaluation}

Planning is triggered by initialization, relevant new evidence, contradictions, execution failures, or an action boundary. The method merges admissible facts, updates each new piece of evidence once, and constructs a finite set of joint search and transport candidates. Every candidate is evaluated on the same batch of sampled worlds. During a rollout, a sampled hidden target becomes available to the deterministic transport planner only after the simulated inspection that would reveal it. Candidate scores include the resulting transport and unfinished-work costs. The selected plan contributes an executable joint action sequence; subsequent real feedback determines the next update.

For candidate $a$ and an available pure-transport baseline $b$, let
$\delta_m=J(a,\omega_m)-J(b,\omega_m)$
be their cost difference in shared world $\omega_m$. A candidate is eligible to replace the baseline only if
\begin{equation}
\overline{\delta}+2\,\mathrm{SE}(\delta)<0.
\label{eq:app-candidate-gate}
\end{equation}
The lowest-mean eligible candidate is selected; if none qualifies, the baseline is retained. Without a pure-transport baseline, candidates within one standard error of the best sampled mean are compared by the analytic gain-to-cost rule. This is a finite-sample decision heuristic, not a statistical significance guarantee. The selected action need not belong to the candidate with the lowest unfiltered mean score.

\begin{table}[t]
\centering
\caption{Planning settings. These are method parameters, separate from the public benchmark configuration.}
\label{tab:app-method-parameters}
\small
\setlength{\tabcolsep}{4pt}
\begin{tabular}{@{}lcc@{}}
\toprule
Parameter & C-WAH & TDW-MAT \\
\midrule
Maximum joint candidates & 6 & 6 \\
Shared sampled worlds & 8 & 8 \\
Rollout event cap & 160 & 160 \\
A* expansion cap in uncertainty evaluation & 2,000 & 2,000 \\
Execution action length & 8 & 5 \\
Completion-delay weight & 0.1 & 0.1 \\
Rollout horizon & 120 estimated seconds & Adaptive path-length scale \\
\bottomrule
\end{tabular}
\end{table}

C-WAH additionally uses a hidden-inspection proxy of 2 and at most two pickup candidates per category. The TDW-MAT implementation uses a $240\times120$ map with 0.125\,m cells, $15^\circ$ turns, and 16 scan orientations. The A* expansion cap in Table~\ref{tab:app-method-parameters} applies to uncertainty evaluation; it does not redefine the deterministic solving path when all relevant targets are known.

\subsection{Planning Costs and Execution Time}
\label{app:planning-cost}

\begin{table}[t]
\centering
\caption{Cost layers used by the method and the evaluator. Planning proxies and recorded EET have different meanings.}
\label{tab:app-cost-layers}
\small
\setlength{\tabcolsep}{3pt}
\renewcommand{\arraystretch}{1.08}
\begin{tabular}{@{}p{.18\columnwidth}p{.38\columnwidth}p{.38\columnwidth}@{}}
\toprule
Layer & C-WAH & TDW-MAT \\
\midrule
Internal A*
& Geometric movement plus unit interaction cost.
& Navigation path length; grasp and put have zero proxy cost. \\
Candidate rollout
& Estimated calibrated seconds, joint synchronization, and estimated message cost.
& Penalized navigation path length and residual unfinished-work cost. \\
Public EET
& Actual displacement, fixed action charges, idle time, and communication critical path.
& Actual joint simulator frames/250, idle time, and communication critical path. \\
\bottomrule
\end{tabular}
\end{table}

Within either rollout, an increment of cost $\Delta\tau$ is charged as
$\Delta J=\Delta\tau(1+0.1u)$,
where $u$ is the unfinished-goal fraction and $\Delta\tau$ retains that rollout's units. C-WAH charges unfinished trajectories over the remainder of its finite 120-second evaluation horizon; insufficient sampled supply cannot end a rollout at zero further cost. Its hidden-search tail is an estimate, not a record of actions executed in the environment. TDW-MAT uses a navigation-scale horizon
$H=\max(1,D)(S+2N+2)$,
where $D$ is the largest finite navigation proxy between public endpoints, $S$ counts scans needed for the next coverage layer, and $N$ counts unfinished targets. This horizon is not measured in seconds. Simulated room scans reveal sampled contents more completely than real local scans may do.

A zero grasp or put proxy in TDW-MAT still changes holdings, object ownership, and completion state. Search-state deduplication includes agent positions and holdings, ongoing actions, object status, and explored state. Zero planning cost therefore does not remove interaction constraints or the simulator frames incurred during real execution. It may, however, favor routes with more interactions; candidate pruning, finite expansion, and feasible fallback add further approximation.

\subsection{Distributed Execution and Scope}
\label{app:method-execution}

C-WAH exports a fact graph from method memory. TDW-MAT separates navigation-facing agent memory from planning-facing method memory. Facts are fused using their source and time information. Agents exchange compact facts, updates, actions, and synchronization information through charged public messages; a simulated message cost never substitutes for actual delivery. Executors manage ongoing actions, role occupancy, and bounded action sequences, while rendezvous and distance recovery rely on public observations and feedback.

The deterministic search operates on its specified state graph and proxy costs. Its conditional shortest-path properties do not imply global optimality in the partially observed two-agent task. Finite candidate sets, sampled worlds, rollout horizons, and execution feedback also prevent a lower planning score from guaranteeing lower real EET. We therefore evaluate the method using recorded task outcomes and execution metrics.
\section{Baseline Adaptation}
\label{app:baselines}

CoELA, CoTS, PCE, and REVECA are connected to the same HiddenCoop task, observation, action, communication, and timing interfaces. We preserve each method's characteristic reasoning pipeline while adapting its execution to hidden cooperation and the public feedback protocol. Because these adaptations can affect trajectories, Table~\ref{tab:app-baseline-versions} distinguishes the retained method core from changes relevant to behavior.

\begin{table}[t]
\centering
\caption{Retained baseline components and behaviorally relevant HiddenCoop adaptations.}
\label{tab:app-baseline-versions}
\small
\setlength{\tabcolsep}{3pt}
\renewcommand{\arraystretch}{1.08}
\begin{tabular}{@{}p{.14\columnwidth}p{.27\columnwidth}p{.52\columnwidth}@{}}
\toprule
Method & Retained core & HiddenCoop adaptation \\
\midrule
CoELA
& Communication and planning calls, object memory, and macro-actions.
& Public goals and carrying capacity are mapped to native memory. Pairing waits are bounded, and reunion responds to public distance feedback. \\
CoTS
& Tree construction, partner review, staged scoring, and action selection.
& In C-WAH, partner review is exchanged through public messages. In TDW-MAT, Alice executes the tree locally with a Bob-role critic; partner state and the final plan are exchanged through public messages. Pending reviews and pairing attempts are bounded; TDW-MAT message compression preserves content and charges all delivered characters. \\
PCE
& Planner--Composer--Evaluator chain, reflection, and utility ordering.
& Public cooperation and capacity information is mapped into the original workflow. Pairing control is released after the specified rejection boundary, and ongoing TDW-MAT actions update public memory. \\
REVECA
& Relevance and proximity retrieval, candidate selection, and trajectory validation.
& Private initial-room sharing is removed. Exact public requests may occupy a candidate slot; cooperative distance considers both agents, and ongoing TDW-MAT steps do not trigger another LLM call. \\
\bottomrule
\end{tabular}
\end{table}

The adapters expose capacity, paired-action requirements, asynchronous continuations, and public pairing, distance, and execution feedback through each method's own decision process. Navigation recovery and rendezvous remain method-side behaviors and are charged under the common action and message rules. No baseline receives complete target locations, hidden cooperation annotations, or \methodname{}'s fused memory.

\subsection{Parameters and Prompt Information}
\label{app:baseline-params}

We retain the method-specific parameter settings and LLM request behavior of the original papers or implementations wherever they remain applicable. The shared model backbone and benchmark interface settings follow the experimental settings. Parameters introduced for pairing, recovery, or interface compatibility are part of the adaptations described above; API fields omitted by an implementation are not assigned inferred values. Usage totals include recorded requests for retries, corrections, or reflection when these occur.

Each baseline builds prompts from the public goals and information available in its own memory, such as local progress, delivered dialogue, action history, and available macro-actions. Cooperative guidance describes how to coordinate on an identified object and how to interpret public feedback; it does not reveal unknown cooperation labels or target locations. A message expressing an intended joint action is distinguished from feedback confirming its execution. These information boundaries apply to both environments despite their different perception and action interfaces.
\section{Evaluation Details}
\label{app:evaluation}

\subsection{Coverage and Recorded Evidence}

In C-WAH, four baselines, \methodname{}, and the single-point system contrast each cover all 100 tasks in three runs, yielding 1,800 episode-run records. The episode seed is $1+100r+i$, where $r$ is the run index and $i$ the task index. TDW-MAT evaluates the four baselines and \methodname{} on all 48 tasks in run 0, yielding 240 episode records. Its experiment seed and task-specific execution seeds are retained in the configurations and records; the task-specific seeds are matched across methods.

The records retain task and source-pair identifiers, task variants, outcomes, stepwise progress, execution-time components, messages, and model usage. Configurations and available source artifacts are indexed with paths and hashes. A hash identifies a stored artifact but cannot recover unrecorded runtime source, hardware, simulator build, or provider-side model defaults. The C-WAH single-point results are treated as a system-version contrast; the five-method main comparison does not include that variant.

\subsection{Metrics and Budget Replay}
\label{app:metric-definitions}

For a group of $N$ valid episodes and budget $B$, task success rate is
$N^{-1}\sum_e\mathbf{1}[t_e^{\mathrm{complete}}\leq B]$.
Goal completion rate pools satisfied target instances over required target instances in the reported group; it is not the unweighted mean of episode-level completion fractions. We distinguish completion by the budget from terminal completion when a retained step crosses $B$. EET and TGC-AUC follow Equations~\ref{eq:app-eet} and~\ref{eq:app-tgc-auc}. Unless a successful-only subset is explicitly named, mean EET includes both valid successes and failures.

Task groups (all, ordinary, and cooperative), outcome groups (all, success, and failure), and target types (ordinary and cooperative) are separate grouping dimensions. A cooperative task may contain ordinary targets. An empty group is reported as unavailable rather than assigned a zero rate.

Input, output, and total LLM tokens and call counts are summed across both agents' recorded attempts, then averaged over episodes. The EET decomposition separates physical time, the idle-only charge, and additional communication time on the joint critical path. LLM and planner wall-clock time are separate diagnostics. In C-WAH, unavailable token fields are excluded from the corresponding field-wise mean. In TDW-MAT, available token usage is summed within each episode, missing call-level fields contribute zero, and the mean uses all valid episodes in the group. Missing-usage calls are recorded separately.

Shorter-budget results replay complete steps from trajectories collected at the main budget: 100--600\,s in 50\,s increments for C-WAH and 10--20\,s in 1\,s increments for TDW-MAT. The first step crossing a shorter budget is retained with its full cost, so terminal EET may exceed $B$; progress first achieved after $B$ does not contribute to budget-limited success or TGC-AUC. These results describe actions of the recorded trajectories, not online reruns in which a method was informed of a different budget.

\subsection{Paired and Conditioned Comparisons}

Ordinary and cooperative variants are matched within each source pair. We report the ordinary-to-cooperative change using paired outcomes, with uncertainty intervals from 5,000 bootstrap resamples of source pairs. C-WAH resamples 50 pairs while keeping each pair's three runs together; TDW-MAT resamples 24 pairs from its single run. These intervals describe variation across the evaluated source pairs and do not quantify independent model or simulator-run variability.

For common-success efficiency, every compared method is evaluated on the same successful episode identifiers. The five main methods share 264 successful C-WAH episode-run records (150 ordinary and 114 cooperative) and 12 successful TDW-MAT episodes (10 ordinary and 2 cooperative). Efficiency comparisons on the two cooperative TDW-MAT episodes are descriptive. Timeout is recorded as a termination condition, not assigned as the sole cause of a failed task.
\section{Extended Results}
\label{app:extended-results}

The analyses below use the five main methods on 100 C-WAH tasks (three runs each) and 48 TDW-MAT tasks (one run each). C-WAH trajectories were collected at 600~s and TDW-MAT trajectories at 20~s. Shorter-budget values replay the recorded trajectories under the strict cutoff rule in Appendix~\ref{app:metric-definitions}; they are not budget-aware reruns. The main paper reports the 200~s C-WAH and 20~s TDW-MAT comparisons.

\subsection{Budgeted Progress}

\begin{figure}[!h]
  \centering
  \includegraphics[width=.49\linewidth]{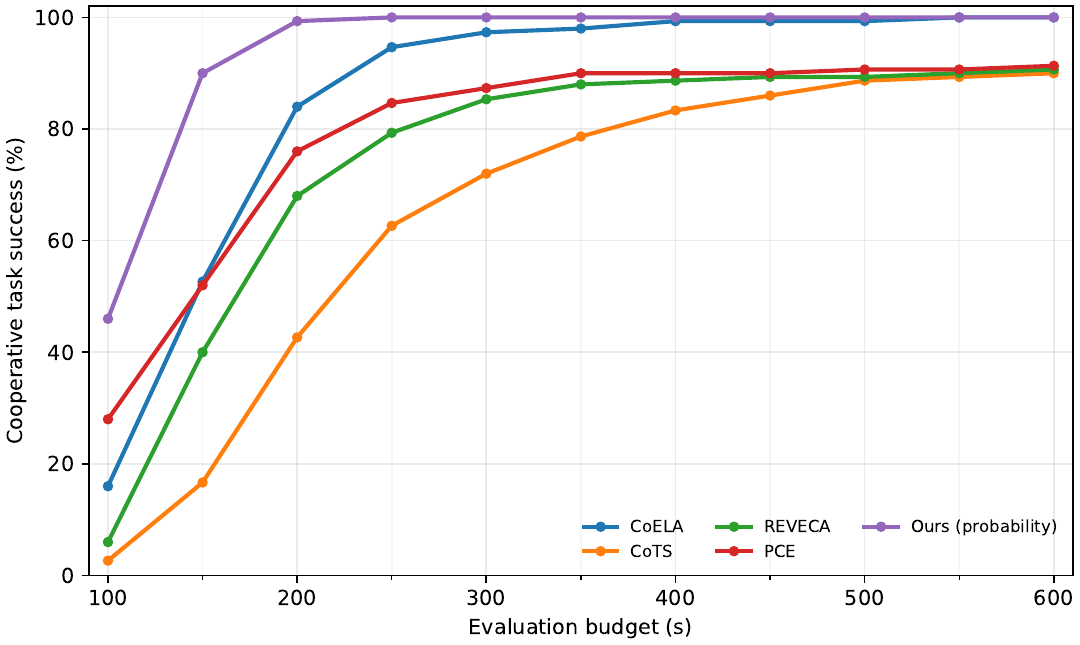}\hfill
  \includegraphics[width=.49\linewidth]{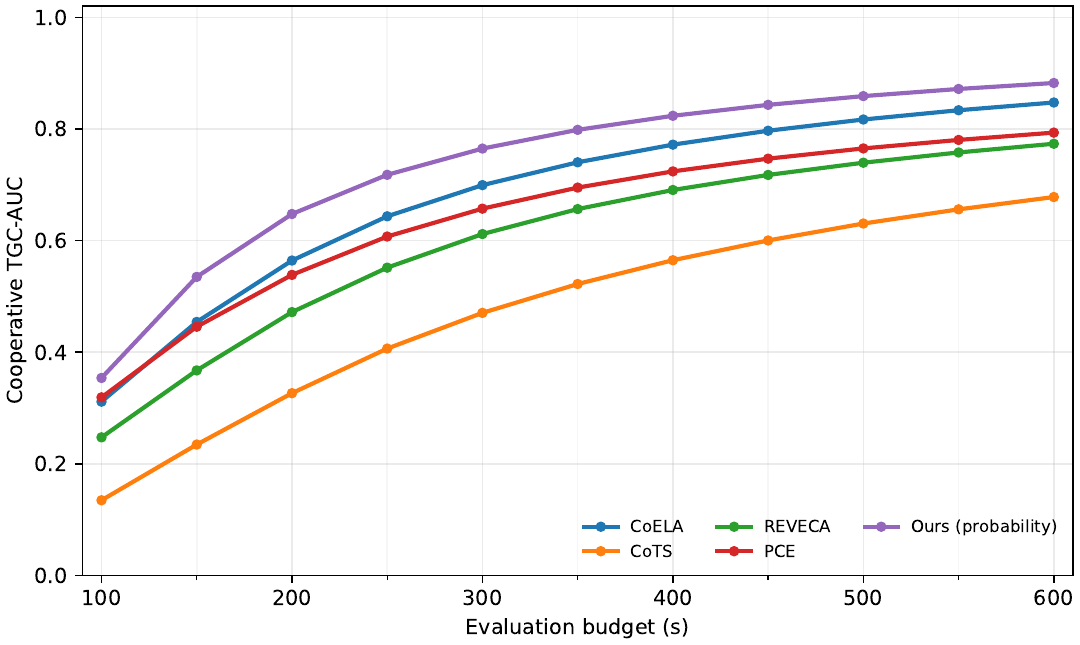}
  \caption{C-WAH cooperative-task success (left) and TGC-AUC (right) under offline budget cutoffs.}
  \label{fig:app-cwah-budget}
\end{figure}

Figure~\ref{fig:app-cwah-budget} separates eventual completion from progress accumulated by a cutoff on cooperative C-WAH tasks. DPGP completes 99.3\% of these records by 200~s and all of them by the evaluated 250~s cutoff. CoELA, the other method with 100\% success at 600~s, first reaches that value at 550~s. The plotted grid does not identify an exact completion threshold between cutoffs.

Table~\ref{tab:app-selected-budgets} gives representative cutoffs for all three task groups in both environments. It complements the cooperative C-WAH curves and makes the TDW-MAT five-method results explicit. On TDW-MAT cooperative tasks, DPGP already succeeds on 33.3\% at 10~s and 58.3\% at 20~s, versus 4.2\% and 45.8\% for the strongest baseline at those cutoffs, PCE. On ordinary TDW-MAT tasks, PCE instead has the higher 20~s success rate (83.3\% versus 70.8\%); the cooperative advantage therefore should not be read as uniform dominance on every task subset.

\begingroup
\small
\setlength{\tabcolsep}{7pt}\renewcommand{\arraystretch}{1.07}
\begin{longtable}{@{}llrrrrr@{}}
\caption{Task success (\%) at selected offline cutoffs. Group labels are shown once.}\label{tab:app-selected-budgets}\\
\toprule
Group & $B$ (s) & CoELA & CoTS & REVECA & PCE & DPGP \\\midrule
\endfirsthead
\toprule
Group & $B$ (s) & CoELA & CoTS & REVECA & PCE & DPGP \\\midrule
\endhead
\bottomrule
\endlastfoot
\multicolumn{7}{@{}l}{\textit{C-WAH (three runs per task)}}\\
All & 200 & 91.7 & 71.0 & 84.0 & 88.0 & 99.7 \\
 & 600 & 100.0 & 95.0 & 95.3 & 95.7 & 100.0 \\\addlinespace[2pt]
Ord. & 200 & 99.3 & 99.3 & 100.0 & 100.0 & 100.0 \\
 & 600 & 100.0 & 100.0 & 100.0 & 100.0 & 100.0 \\\addlinespace[2pt]
Coop. & 200 & 84.0 & 42.7 & 68.0 & 76.0 & 99.3 \\
 & 600 & 100.0 & 90.0 & 90.7 & 91.3 & 100.0 \\\midrule
\multicolumn{7}{@{}l}{\textit{TDW-MAT (one run per task)}}\\
All & 10 & 18.8 & 12.5 & 16.7 & 29.2 & 47.9 \\
 & 15 & 39.6 & 29.2 & 37.5 & 52.1 & 62.5 \\
 & 20 & 43.8 & 47.9 & 52.1 & 64.6 & 64.6 \\\addlinespace[2pt]
Ord. & 10 & 37.5 & 25.0 & 33.3 & 54.2 & 62.5 \\
 & 15 & 58.3 & 54.2 & 54.2 & 70.8 & 70.8 \\
 & 20 & 62.5 & 70.8 & 66.7 & 83.3 & 70.8 \\\addlinespace[2pt]
Coop. & 10 & 0.0 & 0.0 & 0.0 & 4.2 & 33.3 \\
 & 15 & 20.8 & 4.2 & 20.8 & 33.3 & 54.2 \\
 & 20 & 25.0 & 25.0 & 37.5 & 45.8 & 58.3 \\\end{longtable}
\endgroup

\subsection{Paired Cooperation Effects}

Table~\ref{tab:tbl_e_pairs} compares ordinary and cooperative variants within each source pair. Positive entries denote a cooperation-induced loss in success or AUC, or an EET increase. The 95\% intervals resample 50 C-WAH source pairs with their repeated runs kept together, or 24 TDW-MAT pairs from the single run. DPGP has the smallest AUC loss and EET increase in both environments. On TDW-MAT its success-rate loss is also the smallest, although the interval includes zero.

\par\smallskip\noindent
\begin{minipage}{\linewidth}
\captionof{table}{Ordinary-to-cooperative paired differences with source-pair bootstrap 95\% intervals. SR losses are percentage points.}
\label{tab:tbl_e_pairs}
\vspace{3pt}
\centering\footnotesize
\renewcommand{\arraystretch}{1.07}
\begin{tabular*}{\linewidth}{@{\extracolsep{\fill}}lrrr@{}}\toprule
Method & SR loss & AUC loss & $\Delta$EET (s) \\\midrule
\multicolumn{4}{@{}l}{\textit{C-WAH}}\\
CoELA & 0.0 [0.0, 0.0] & .0488 [.0376, .0615] & 46.59 [37.50, 56.52] \\
CoTS & 10.0 [5.3, 15.3] & .1851 [.1438, .2273] & 141.70 [115.10, 170.22] \\
REVECA & 9.3 [4.7, 14.7] & .1353 [.1007, .1752] & 121.41 [95.22, 151.38] \\
PCE & 8.7 [2.7, 16.0] & .1200 [.0699, .1770] & 97.25 [64.44, 136.69] \\
DPGP & 0.0 [0.0, 0.0] & .0265 [.0211, .0322] & 25.46 [19.30, 32.13] \\\midrule
\multicolumn{4}{@{}l}{\textit{TDW-MAT}}\\
CoELA & 37.5 [20.8, 58.3] & .2183 [.1710, .2711] & 4.91 [2.90, 7.02] \\
CoTS & 45.8 [20.8, 66.7] & .2181 [.1632, .2780] & 4.95 [3.17, 6.79] \\
REVECA & 29.2 [4.2, 54.2] & .1738 [.0870, .2636] & 4.61 [2.59, 6.61] \\
PCE & 37.5 [16.7, 58.3] & .1701 [.1077, .2386] & 5.72 [4.06, 7.39] \\
DPGP & 12.5 [0.0, 25.0] & .1099 [.0605, .1733] & 3.68 [2.25, 5.27] \\\bottomrule
\end{tabular*}
\end{minipage}
\par\smallskip

\subsection{Common-Success Efficiency and Target Types}

A success-conditioned mean can favor a method that only completes easy tasks. Table~\ref{tab:tbl_e_common} instead uses the same successful episode identifiers for every method. DPGP has the lowest EET and highest AUC on the 264-record C-WAH intersection, including its 114 cooperative records, and on the 12-record TDW-MAT intersection. Only two TDW-MAT cooperative episodes belong to the latter intersection, so we do not rank methods on that subset separately.

\par\smallskip\noindent
\begin{minipage}{\linewidth}
\captionof{table}{Five-method common-success efficiency. EET is in seconds; higher TGC-AUC is better.}
\label{tab:tbl_e_common}
\vspace{3pt}
\centering\footnotesize
\renewcommand{\arraystretch}{1.07}
\begin{tabular*}{\linewidth}{@{\extracolsep{\fill}}lrrrrrr@{}}\toprule
 & \multicolumn{2}{c}{C-WAH all ($N=264$)} & \multicolumn{2}{c}{C-WAH coop. ($N=114$)} & \multicolumn{2}{c}{TDW-MAT all ($N=12$)}\\
Method & EET & AUC & EET & AUC & EET & AUC \\\midrule
CoELA & 123.97 & .8779 & 146.59 & .8537 & 10.20 & .6826 \\
CoTS & 165.85 & .8176 & 219.07 & .7576 & 11.78 & .6532 \\
REVECA & 122.84 & .8766 & 163.07 & .8342 & 11.74 & .6543 \\
PCE & 113.14 & .8866 & 143.49 & .8510 & 9.87 & .6965 \\
DPGP & \textbf{90.25} & \textbf{.9003} & \textbf{100.70} & \textbf{.8888} & \textbf{7.04} & \textbf{.7725} \\\bottomrule
\end{tabular*}
\end{minipage}
\par\smallskip

Cooperative episodes include both ordinary and cooperation-requiring targets. Table~\ref{tab:tbl_e_goals} separates these target types and pools terminal completed/required instances, rather than averaging task-level percentages. TDW-MAT DPGP completes the most targets overall (82/103), while PCE completes one more cooperation-requiring target (34/46 versus 33/46); task success and target completion therefore capture distinct aspects of performance.

\par\smallskip\noindent
\begin{minipage}{\linewidth}
\captionof{table}{Terminal target completion in cooperative episodes (completed/required); C-WAH pools three runs, TDW-MAT one.}
\label{tab:tbl_e_goals}
\vspace{2pt}
\centering\small
\renewcommand{\arraystretch}{1.06}
\begin{tabular*}{\linewidth}{@{\extracolsep{\fill}}lrrrrrr@{}}\toprule
 & \multicolumn{3}{c}{C-WAH} & \multicolumn{3}{c}{TDW-MAT}\\
Method & All & Ordinary & Coop. & All & Ordinary & Coop. \\\midrule
CoELA & 627/627 & 285/285 & 342/342 & 69/103 & 41/57 & 28/46 \\
CoTS & 578/627 & 272/285 & 306/342 & 68/103 & 41/57 & 27/46 \\
REVECA & 592/627 & 283/285 & 309/342 & 72/103 & 46/57 & 26/46 \\
PCE & 595/627 & 285/285 & 310/342 & 80/103 & 46/57 & 34/46 \\
DPGP & 627/627 & 285/285 & 342/342 & 82/103 & 49/57 & 33/46 \\\bottomrule
\end{tabular*}
\end{minipage}
\par\smallskip

\subsection{Execution and Model Cost}

Across three C-WAH runs, DPGP succeeds on all 100 tasks each time (EET 93.92--96.32~s; AUC .8956--.8960). Table~\ref{tab:tbl_e_costs} reports all-episode means at the collection budgets. Physical time and the additional communication charge contribute to EET; tokens and calls are separate model-use diagnostics. DPGP uses fewer tokens and calls than every baseline in both environments, although its message count and communication charge are not uniformly smallest.

\par\smallskip\noindent
\begin{minipage}{\linewidth}
\captionof{table}{All-episode execution and model-use means at the collection budgets. Tokens are thousands per episode.}
\label{tab:tbl_e_costs}
\vspace{3pt}
\centering\footnotesize
\renewcommand{\arraystretch}{1.06}
\begin{tabular*}{\linewidth}{@{\extracolsep{\fill}}lrrrrr@{}}\toprule
Method & Physical (s) & Comm. extra (s) & Messages & Tokens (k) & Calls \\\midrule
\multicolumn{6}{@{}l}{\textit{C-WAH (600~s)}}\\
CoELA & 129.77 & .300 & 23.5 & 143.56 & 128.0 \\
CoTS & 162.87 & 33.385 & 42.0 & 162.69 & 125.7 \\
REVECA & 149.11 & 2.300 & 24.9 & 287.17 & 199.6 \\
PCE & 138.69 & .001 & 1.9 & 123.12 & 72.5 \\
DPGP & 87.85 & 7.177 & 30.6 & 7.16 & 2.6 \\\midrule
\multicolumn{6}{@{}l}{\textit{TDW-MAT (20~s)}}\\
CoELA & 15.78 & .000 & 4.3 & 485.84 & 501.8 \\
CoTS & 14.35 & 2.329 & 443.2 & 2825.56 & 1327.7 \\
REVECA & 15.49 & .109 & 74.9 & 1013.29 & 804.9 \\
PCE & 14.22 & .013 & 27.2 & 1202.03 & 635.5 \\
DPGP & 11.71 & .590 & 98.8 & 1.41 & 2.0 \\\bottomrule
\end{tabular*}
\end{minipage}
\par\smallskip

\subsection{Probability and Single-Point System Contrast}
\label{app:system-contrast}

The C-WAH single-point system assigns definite location hypotheses before deterministic joint search. Table~\ref{tab:tbl_e_versions} compares it with the probability version at an early cutoff and at the collection budget. Both systems eventually complete all 300 records, but the probability version finishes sooner and accumulates more progress.

\par\smallskip\noindent
\begin{minipage}{\linewidth}
\captionof{table}{C-WAH probability/single-point comparison (P/D); EET in seconds. Each task group is shown once.}
\label{tab:tbl_e_versions}
\vspace{3pt}
\centering\small
\renewcommand{\arraystretch}{1.06}
\begin{tabular*}{\linewidth}{@{\extracolsep{\fill}}llrrr@{}}\toprule
Group & $B$ (s) & SR (\%) P/D & EET P/D & AUC P/D \\\midrule
All & 150 & 95.0/69.7 & 93.61/126.88 & .5856/.4738 \\
 & 600 & 100.0/100.0 & 95.04/133.98 & .8957/.8649 \\\addlinespace[2pt]
Ord. & 150 & 100.0/85.3 & 82.31/119.49 & .6360/.5129 \\
 & 600 & 100.0/100.0 & 82.31/120.92 & .9090/.8776 \\\addlinespace[2pt]
Coop. & 150 & 90.0/54.0 & 104.90/134.28 & .5351/.4347 \\
 & 600 & 100.0/100.0 & 107.76/147.04 & .8825/.8521 \\\bottomrule
\end{tabular*}
\end{minipage}
\par\smallskip

\section{Trace-Level Case Study}
\label{app:case-study}

We use C-WAH task16 to illustrate how belief revision changes executable exploration. The ordinary task requires one apple, one wine, and one pudding on coffee table 268. The probability and single-point versions start from the same scene, but their internal decisions and subsequent observations need not remain identical; the case is illustrative rather than a controlled module ablation.

At the first probability decision, support levels 0 and 1 for zero versus one anonymous object normalize to probabilities $1/3$ and $2/3$. Six candidates receive mean scores 122.85, 123.16, 122.30, 122.13, 125.98, and 121.90. Candidate 0 is selected even though candidate 5 has the smallest displayed mean, consistent with the recorded one-standard-error eligibility and analytic tie break.

When a new container appears at step 6, the probability version assigns one agent to continued exploration and the other to dishwasher 104; the single-point version instead schedules an unknown-region alternative in bathroom 172. The probability trace opens the dishwasher at step 13 and records wine 368 and apple 373 by step 15, then opens the microwave at step 20 and grasps pudding 370 at step 23. The single-point trace reaches the relevant refrigerator/microwave stage around step 51. These are control-step indices, not uniform time units.

\begin{figure}[tbp]
\centering
\includegraphics[width=.82\linewidth]{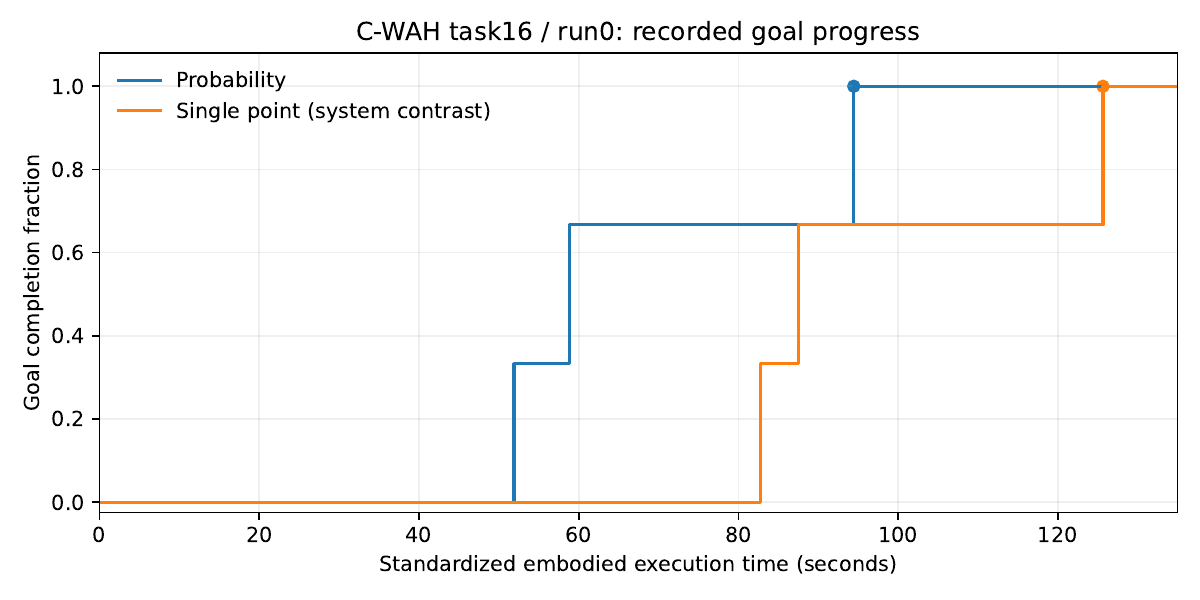}
\caption{Task16 goal progress against recorded embodied execution time. Both versions succeed.}
\label{fig:app-task16-progress}
\end{figure}

The probability trace delivers pudding, wine, and apple at steps 46, 51, and 84, completing in 94.4532\,s with AUC 0.6579. The single-point trace delivers wine, pudding, and apple at steps 73, 76, and 111, completing in 125.6664\,s with AUC 0.5067. Their recorded model usage is 2 calls/5,672 tokens versus 12 calls/17,174 tokens; neither trace records an execution-failure event. The 31.2132\,s difference is consistent with the earlier discovery schedule, but cannot be attributed solely to uncertainty representation because the system versions are not otherwise identical. As an ordinary episode, task16 demonstrates search and transport scheduling rather than the paired-manipulation transaction.

\section{LLM Prompts}
\label{app:prompts}
The only use of large language models (LLMs) in our method is the semantic
discovery-support estimation described in Appendix~\ref{app:belief}; all other components
are deterministic. At each request, the method assembles the required rows
(anonymous count hypotheses or semantic location groups) from public task
information, and the LLM returns an integer support level $L_j \in \{0,\dots,4\}$
for every candidate, which is converted into relative weights by Equation~\eqref{eq:app-support-normalization}. The prompt structure is shared across agents and requests; the
rows supplied as input differ. The complete C-WAH and TDW-MAT prompts are listed below.

\definecolor{cwahbg}{RGB}{241,246,255}
\definecolor{cwahframe}{RGB}{68,133,199}

\definecolor{tdwbg}{RGB}{255,232,212}
\definecolor{tdwframe}{RGB}{212,86,46}

\begin{promptbox}
  {Discovery-support estimation prompt (C-WAH)}
  {cwahbg}
  {cwahframe}
\begin{lstlisting}[language=prompt]
Estimate a finite discovery model from public household semantics.

OUTPUT FORMAT:

Return only a json object with exactly one field, "rows", containing an array. Each row has EXACTLY "key" and "levels". Copy its key and every candidate_id from required_rows exactly once. Every value in "levels" must be one of the JSON numbers 0, 1, 2, 3 or 4. Never return probabilities, counts, arrays, input metadata or explanations in a row.

Fictional format example (NOT this request): {"rows":[{"key":"count:fictional_item","levels":{"n0":1,"n1":3,"n2":2,"n3":0}}, {"key":"place:fictional_item","levels":{"g0":1,"g1":4}}]}. Notice that n3 identifies the capped count hypothesis 3, while its value 0 is a support level. Never copy the example keys, candidate IDs or levels into your answer. Return all required rows exactly once in the supplied order. No markdown or extra fields.

MEANING OF LEVELS:

Levels mean relative support: 0=very unlikely but possible, 1=unlikely, 2=neutral/insufficient evidence, 3=plausible, 4=strongly plausible. These are NOT probabilities. Code maps each level L to 2**L and normalizes the row.

Equal levels are valid when evidence does not distinguish alternatives. Do not force differences or assign high support simply because a task needs an object. Judge whether a useful object is physically stored there now and would be found by inspection.

This is storage/discovery likelihood, not a broad association between the object and activities performed near that location. Consider the object class, whether the location normally stores or contains such loose objects, and its room together. Room context may adjust but must not override an unsuitable location function. An unexplored location means missing evidence, not a semantic match.

Level 2 means the supplied evidence truly does not distinguish the alternatives; use lower support when the location's normal storage purpose clearly differs. Generic example: clothing is usually more discoverable in bedroom storage than in a kitchen refrigerator, although an atypical location remains possible.

For kind="count_support", useful_limit is the number of unfinished task slots not already covered by known identities. Each candidate explicitly gives an id, count and relation. relation="exact" means exactly that many useful anonymous objects; relation="at_least" is the final saturated hypothesis and means enough anonymous objects to cover useful_limit, possibly with irrelevant extras. For useful_limit=1, judge only none versus at least one. For useful_limit=3, judge 0, 1, 2 and at least 3. Extra objects cannot fill extra task slots. Copy the n-prefixed id; never use the count itself as the support level. Do NOT sum levels to one.

For kind="weights", candidates explicitly describe semantic location groups. Return one levels entry for each candidate id. Containers of the same class in the same room share a group: rate the TOTAL allocation mass of that group. Code normalizes group weights, then divides each group's mass equally among its real members. More copies of a container are not, by themselves, stronger evidence.

Visible areas, hidden areas and unmapped residuals remain separate alternatives. Distinguish groups when household storage semantics supports a difference; use neutral levels when it does not. Do not give high support to a fixture merely because an object is used in the same room; ask whether inspecting it would normally reveal the object inside. A location can be atypical without being impossible. Rate relative discovery plausibility. All-zero levels are valid neutral relative support. Different objects may occupy the same region. Do not return a location ID.

When newly observed containers change the candidates, compare every supplied group again; a new container's semantics may change its room's support relative to other rooms. Do not preserve an earlier room total merely because it was predicted first. For role="goal_target", locate the furniture itself, not the objects to deliver.

PUBLIC EVIDENCE:

Exclude every known identity from anonymous counts, including held or unlocated known objects. Task demand is not a count of hidden supply; known objects are not new discoveries. Task destinations are not evidence of initial object locations.

The candidate fields room, location and kind state the relevant public semantics. A checked visible area does not mean that closed containers there are empty. Use the supplied public semantics only; group IDs are labels, not semantic clues.

Household semantics are a soft prior: atypical locations remain possible, and repeated containers do not imply proportionally more hidden supply. Estimate anonymous supply separately from task demand; do not assume every inspected container yields a useful object. Do not predict cooperative labels, navigation costs or actions.

Before responding, check the row count, exact field names, every candidate ID, levels 0..4, and absence of extra fields.
\end{lstlisting}
\end{promptbox}

\begin{promptbox}
  {Discovery-support estimation prompt (TDW-MAT)}
  {tdwbg}
  {tdwframe}
\begin{lstlisting}[language=prompt]
Estimate where still-unlocated household task objects are physically likely to be stored and discovered, using household commonsense and the supplied public facts. Judge each concrete object class and each room's function together. Both a relevant functional room and a shared living space can be plausible. A broad association with an activity is weaker evidence than a room's normal role in storing or keeping that object. Personal belongings need not have the same distribution as specialized work items.

For every query, rate EVERY supplied room candidate: 0=very unlikely but possible, 1=unlikely, 2=neutral/insufficient evidence, 3=plausible, 4=strongly plausible. These are relative support levels, NOT probabilities; code normalizes 2**level. Distinguish rooms when semantics or public discoveries justify it. Equal levels are appropriate when evidence does not distinguish alternatives, including equally plausible rooms with the same function. Room IDs and candidate keys are labels, not semantic clues.

First-discovery locations are positive evidence, not exclusive class locations. Do not infer initial storage from the task destination or transported objects. Unexplored means missing evidence, not high support. A visual scan can miss an object: no empty scan proves absence. Atypical rooms remain possible. Do not invent observations, objects, rooms, coordinates or cooperative properties. Do not choose actions or explain the answer.

Return each query_key and every candidate_key exactly once, with integer levels 0..4 and no extra fields. Return only JSON in this exact shape: {"rows":[{"query_key":"place:fictional_item","levels":{"g0":2,"g1":4}}]} The example is structural only; never copy its labels or levels.
\end{lstlisting}
\end{promptbox}

\end{document}